\documentclass[a4paper,11pt]{article}
\usepackage{pos}

\newcommand{\be}{\begin{equation}}
\newcommand{\ee}{\end{equation}}
\newcommand{\ba}{\begin{eqnarray}}
\newcommand{\ea}{\end{eqnarray}}
\newcommand{\ban}{\begin{eqnarray*}} 
\newcommand{\ean}{\end{eqnarray*}}
\newcommand \nn {\nonumber}

\def\P{{\mathbf P}}
\def\b{{\mathbf b}}

\def\k{{\mathbf k}}

\def\z{{\mathbf z}}

\title{Correlation limit of DIS dijet production at the subeikonal order and gluon TMDs}

\author*[a]{Alina Czajka}
\author[a]{Tolga Altinoluk}
\author[a]{Guillaume Beuf}
\author[b]{Cyrille Marquet}

\affiliation[a]{Theoretical Physics Division, National Centre for Nuclear Research,\\
Pasteura 7, Warsaw 02-093, Poland}

\affiliation[b]{CPHT, CNRS, \'Ecole polytechnique, Institut Polytechnique de Paris,\\ 91120 Palaiseau, France}

\emailAdd{tolga.altinoluk@ncbj.gov.pl}
\emailAdd{guillaume.beuf@ncbj.gov.pl}
\emailAdd{alina.czajka@ncbj.gov.pl}
\emailAdd{cyrille.marquet@polytechnique.edu}

\abstract{In these proceedings we present a concise summary of results on the dijet production in Deep Inelastic Scattering (DIS) at small x at the next-to-eikonal accuracy. The presented results were obtained in the correlation limit of the produced jets, which provides a convenient framework to study the emerging transverse-momentum-dependent (TMD) correlators of gluon fields. The dijet production cross section is therefore written in the form of higher-twist expansion in terms of the gluon-field correlators, which are further related to the gluon TMDs distributions. In particular, we show how the $x$ dependent phase of the twist-2 and twist-3 gluon TMDs emerges from the subeikonal corrections calculated within Color Glass Condensate (CGC) framework.}

\FullConference{13th International Conference on Hard and Electromagnetic Probes of High-Energy Nuclear Collisions (HardProbes2026)\\
21-26 June 2026\\
Nashville, Tennessee, USA\\}

\begin{document}
\maketitle

\section{Introduction}

Forward dijet or dihadron production in DIS has been proposed as one of the main processes to study gluon saturation at low x, see Refs.~\cite{Marquet:2007vb,Zheng:2014vka}. However, it was later understood that Sudakov effects also suppress the back-to-back production in this context, like gluon saturation effects. This motivates in-depth studies of the interplay between gluon saturation at low x and Sudakov suppression, notably in the context of the upcoming era of Electron-Ion Collider (EIC). In particular, due to the moderate energies of the EIC, subeikonal corrections to the low-x formalism (power-suppressed in energy) should be understood in order to reach sufficient precision. Over last years, a significant progress in extracting physics behind subeikonal effects has been achieved, see, for example \cite{Altinoluk:2014oxa,Altinoluk:2015gia,Altinoluk:2015xuy,Agostini:2019avp,Agostini:2019hkj,Agostini:2022ctk,Agostini:2022oge,Altinoluk:2020oyd,Altinoluk:2021lvu,Altinoluk:2022jkk,Agostini:2024xqs,Altinoluk:2023qfr}. Among many pursued directions is the importance of subeikonal corrections in studies on the connection between the CGC framework and the TMDs factorization framework. 

The interplay between the CGC and TMD frameworks beyond eikonal accuracy is most conveniently studied in the correlation limit, where the two produced jets scatter back-to-back in the CGC medium, see \cite{Dominguez:2011wm,Kotko:2015ura,Marquet:2016cgx,Altinoluk:2019fui,Altinoluk:2019wyu,Boussarie:2020vzf,Mukherjee:2026cte}. The whole system is then characterized by two momentum scales: ${\bf k}$ being the total momentum of the jets and $\bf P$ being the relative momentum between the jets. The correlation limit holds for $|{\bf k}| \ll |{\bf P}|$, which is also the origin of the higher-twist expansion with the corrections of the type $(|{\bf k}|/|{\bf P}|)^n$ or the corresponding scaling of the color operators.

In this work we shortly summarize the main results of the analysis of the DIS quark-antiquark pair production at the next-to-eikonal accuracy in the correlation limit, which is comprehensively discussed in our original work, Ref.~\cite{Altinoluk:2024zom}. Here we only discuss the results obtained for the longitudinally polarized photon. The analysis here focuses only on the case where quark and antiquark of the dijet interact with gluon background field. A similar study, where the interaction with background quark and antiquark fields is included, is shown in Ref.~\cite{Altinoluk:2026dmt}.

\section{Results}

The DIS dijet production cross section at the next-to-eikonal accuracy is obtained by squaring the respective scattering amplitude. The analysis of the process, derivation of the amplitude, notation, and discussion of the contributing terms is discussed in detail in Ref.~\cite{Altinoluk:2024zom}. For derivation of the propagators at the next-to-eikonal accuracy and discussion of the process in general kinematics see Refs.~\cite{Altinoluk:2020oyd,Altinoluk:2022jkk}. 

The eikonal contribution to the cross section is obtained by squaring the eikonal amplitude and the subeikonal corrections are found as the interference terms between the eikonal part of the amplitude and the next-to-eikonal terms. The resulting expression can be next reorganized so that to group terms with respect to emerging correlators of the gluon-field components. Thus, the cross section at the subeikonal accuracy can be, in general, written as
%
\begin{align}
&
\frac{d\sigma_{\gamma^{*}_L\rightarrow q_1\bar q_2}}{d {\rm P.S.}}\Bigg|_{{\rm Eik }+{\rm NEik}}
= \frac{d\sigma_{\gamma^{*}_L\rightarrow q_1\bar q_2}}{d {\rm P.S.}}\Bigg|_{{\mathcal{F}^{\perp -}} {\mathcal{F}^{\perp -}}}
+\frac{d\sigma_{\gamma^{*}_L\rightarrow q_1\bar q_2}}{d {\rm P.S.}}\Bigg|_{{\mathcal{F}^{\perp -}} {\mathcal{F}^{\perp -}} {\mathcal{F}^{\perp -}}}
+ \frac{d\sigma_{\gamma^{*}_L\rightarrow q_1\bar q_2}}{d {\rm P.S.}}\Bigg|_{{\mathcal{F}^{+ -}} {\mathcal{F}^{\perp -}}}
\, ,
\label{X_sec_L_gen_form_a}
\end{align}
%
where the consecutive terms correspond to contributions involving 2-body gluon-field $\mathcal{F}^{\perp -}$ correlators, 3-body gluon-field $\mathcal{F}^{\perp -}$ correlators, and 2-body correlators with subleading contribution from the $\mathcal{F}^{+-}$ component, respectively.
The three consecutive contributions in the expression~(\ref{X_sec_L_gen_form_a}) are given by
%
%
\begin{align}
&\, \frac{d\sigma_{\gamma^{*}_L\rightarrow q_1\bar q_2}}{d {\rm P.S.}}\Bigg|_{{\mathcal{F}^{\perp -}} {\mathcal{F}^{\perp -}}}
= 2q^+ 2\pi\,  \delta(q^+\!-\!k^+) \, 
\,  4\, Q^2\,  e^2 e_f^2\, z_1^3 z_2^3
\Bigg[
\frac{4\P^i\P^j}{[\P^2\!+\!\bar Q^2]^4}
-2(z_2\!-\!z_1)\frac{(\P^i\k^j+\k^i\P^j)}{[\P^2\!+\!\bar Q^2]^4}
\nonumber\\
 &
 +16(z_2\!-\!z_1)\frac{(\k\!\cdot\!\P)\P^i\P^j}{[\P^2\!+\!\bar Q^2]^5}
\Bigg]
  \int_{\b,\b'}\,   e^{i(\b'-\b)\cdot\k}\, 
   \int  dv^+ d{v'}^+\,
 \bigg[ 1-
 i\frac{[\P^2\!+\!\bar Q^2]}{2q^+z_1 z_2} \, 
 ({v'}^+\!-\!v^+)
 \bigg]
\nonumber\\
&
\times\, 
 \Big\langle 
 g{\mathcal{F}_{j}^{\;-}}_b ({v'}^+,\b')
 \Big[  \mathcal{U}_A\left(+\infty,{v'}^+;\b'\right)^{\dag}  \mathcal{U}_A\left(+\infty,v^+;\b\right)\Big]_{ba}
  g{\mathcal{F}_{i}^{\;-}}_a (v^+,\b)\Big\rangle
 +O\left(\frac{Q^2}{\P^8}\right)
 +O\left(\frac{Q^2}{\k^2 \P^2 W^4}\right)
\label{X_sec_L_Fperpmin_Fperpmin}
\end{align}
%
and
%
%
\begin{align}
&
\frac{d\sigma_{\gamma^{*}_L\rightarrow q_1\bar q_2}}{d {\rm P.S.}}\Bigg|_{{\mathcal{F}^{\perp -}} {\mathcal{F}^{\perp -}} {\mathcal{F}^{\perp -}}}
= 2q^+ 2\pi\,  \delta(q^+-k^+) \, 16\, Q^2\,  e^2 e_f^2\, z_1^3 z_2^3\, \frac{\P^l}{[\P^2\!+\!\bar Q^2]^4}\, 
\nn \\
&    
2\, {\textrm{Re}}\;    
\,  
 \int_{\b,\b'}\,   e^{i(\b'-\b)\cdot\k}\, 
  \int  dv^+ dw^+ d{v'}^+
 {\rm tr}_F\big( t^{a'} t^{b'} t^{c'}\big)\;
 \Big\langle   \mathcal{U}_A\left(+\infty,{v'}^+;\b'\right)_{c'c}  g{\mathcal{F}_{l}^{\;-}}_c ({v'}^+,\b')
 \nn \\
 &
 \times
 \mathcal{U}_A\left(+\infty,{v}^+;\b\right)_{a'a}  g{\mathcal{F}_{i}^{\;-}}_a ({v}^+,\b)
  \mathcal{U}_A\left(+\infty,{w}^+;\b\right)_{b'b}  g{\mathcal{F}_{j}^{\;-}}_b ({w}^+,\b)
  \Big\rangle
   \nonumber\\
 &
 \times
 \Bigg\{
\delta^{ij}\bigg[1-  i\frac{[\P^2\!+\!\bar Q^2]}{2q^+z_1 z_2} \, 
 \left({v'}^+\!-\!\min(v^+,w^+)\right)\bigg]
-\frac{4\, \P^i\P^j}{[\P^2\!+\!\bar Q^2]}
\bigg[1-  i\frac{[\P^2\!+\!\bar Q^2]}{2q^+z_1 z_2} \, 
 \left({v'}^+\!-\!z_2 v^+\!-\!z_1 w^+)\right)\bigg]
   \Bigg\}
  \nonumber\\
 &  
   +O\left(\frac{Q^2}{\P^8}\right)
 +O\left(\frac{Q^2}{\k^2 \P^2 W^4}\right)
   \, ,
\label{X_sec_L_Fperpmin_Fperpmin_Fperpmin}   
\end{align}
and
\begin{align}
&\, \frac{d\sigma_{\gamma^{*}_L\rightarrow q_1\bar q_2}}{d {\rm P.S.}}\Bigg|_{{\mathcal{F}^{+ -}} {\mathcal{F}^{\perp -}}}
=  2q^+ \, 2\pi\, \delta(q^+-k^+) 
\,  8\, Q^2\,  e^2 e_f^2\, \frac{z_1^2z_2^2 (z_2\!-\!z_1)}{q^+} \; \frac{\P^j[\P^2+m^2]}{[\P^2\!+\!\bar Q^2]^4}\,
 2\, {\textrm{Re}}\,  \int_{\b,\b'}\,   e^{i(\b'-\b)\cdot\k}\,
\nn \\
&
\times
 \int  dv^+ d{v'}^+
 \Big\langle 
 g{\mathcal{F}_{j}^{\;-}}_b ({v'}^+,\b')
 \Big[  \mathcal{U}_A\left(+\infty,{v'}^+;\b'\right)^{\dag}  \mathcal{U}_A\left(+\infty,v^+;\b\right)\Big]_{ba}
g\mathcal{F}^{\, +-}_a (v^+,\b)\Big\rangle
\nn \\
 &
+O\left(\frac{Q^2}{\P^6 W^2}\right)
+O\left(\frac{Q^2}{|\k|\, |\P|^3 W^4}\right)
  \, .
\label{X_sec_L_Fplusmin_Fperpmin}   
\end{align}
%
%
%
%

Note that the expressions~\eqref{X_sec_L_Fperpmin_Fperpmin} and~\eqref{X_sec_L_Fperpmin_Fperpmin_Fperpmin} both contain eikonal and next-to-eikonal contributions and Eq.~\eqref{X_sec_L_Fplusmin_Fperpmin} is of the subeikonal order. An important observation in these results is that the next-to-leading-power (NLP) and the next-to-eikonal power corrections factorize from each other in Eq.~\eqref{X_sec_L_Fperpmin_Fperpmin}. In that expression, the next-to-leading-power corrections are purely kinematical corrections to the hard factor, while the next-to-eikonal correction (involving the interval ${v'}^+\!-\!v^+$ between the two field strength insertions) is a correction to the operator. The 3-body term~\eqref{X_sec_L_Fperpmin_Fperpmin_Fperpmin} contains 3-body genuine twists corrections, which are power suppressed but eikonal, and other terms that are both NLP and next-to-eikonal. 

To connect the cross section given by the expression~(\ref{X_sec_L_gen_form_a}) with the TMD results we use the following notation for the gluon field strength correlator in an unpolarized target
%
\begin{align}
\Phi^{\mu\nu;\rho\sigma}({\rm x},\k) \equiv&\,
\frac{1}{{\rm x} P_{{tar}}^-}\, \frac{1}{(2\pi)^3}
 \int d^2\z\,   e^{-i\k \cdot\z}\, 
   \int  d z^+\, e^{i {\rm x} P_{{tar}}^- z^+}\, 
\nonumber\\
 &\, \times\, 
  \Big\langle P_{{tar}}\Big|
 \mathcal{F}^{\mu\nu}_b (0)
 \Big[  \mathcal{U}_A\left(+\infty,0;0\right)^\dagger  \mathcal{U}_A\left(+\infty,z^+;\z\right)\Big]_{ba}
  \mathcal{F}^{\rho\sigma}_a (z)\Big| P_{{tar}}\Big\rangle \bigg|_{z^-=0}
\label{def_Phi_correlator_Fmunu_Frhosigma}
\, ,
\end{align}
with a gauge link forming a future pointing staple. The angle brackets in the correlator (\ref{def_Phi_correlator_Fmunu_Frhosigma}) should be understood as
\ba
\langle{\cal O} \rangle &=& \lim_{P_{{tar}}'\rightarrow P_{{tar}}}
\frac{\langle P_{{tar}}'|\hat{\cal O}| P_{{tar}}\rangle}{\langle P_{{tar}}'| P_{{tar}}\rangle}
\, ,
\label{def_average}
\ea
that is, they denote the expectation value of an operator in the quantum state of the target
(see, for example, Refs.~\cite{Dominguez:2011wm,Altinoluk:2019wyu}). The target states are normalized as follows
\ba
\langle P_{{tar}}'| P_{{tar}}\rangle &=& 2P_{{tar}}^-\, (2\pi)^3 \delta(P_{{tar}}^{'-}\!-\!P_{{tar}}^-)\, \delta^{(2)}(\P_{{tar}}'\!-\!\P_{{tar}})
\, .
\label{norm_states}
\ea

Using the formula (\ref{def_Phi_correlator_Fmunu_Frhosigma}), the cross section in Eq.~\eqref{X_sec_L_gen_form_a} can be rewritten as
%
\begin{align}
\frac{d\sigma_{\gamma^{*}_{L}\rightarrow q_1\bar q_2}}{d {\rm P.S.}}\Bigg|_{{\rm Eik }+{\rm NEik}} \!\!\!
= &\,
2q^+ \, 2\pi\, \delta(q^+\!-\!k^+) 
\,  z_1 z_2\,  e^2 e_f^2\, g^2 \; \frac{(2\pi)^3}{2}\,
\nn \\
& \,
\times
\Bigg\{
{\cal C}^{ij}_{L}(z_1,\P,\k)
 \bigg[ 1 +\frac{[\P^2\!+\!\bar Q^2]}{z_1 z_2(2q^+P_{{tar}}^-)} \, \partial_{{\rm x}}\bigg]  
 \bigg[ {\rm x}\, \Phi^{j -;i-}({\rm x},\k) \bigg] 
 \nn\\
 &\,
 -\frac{1}{2q^+}\, {\cal C}^{j}_{L}(z_1,\P)\bigg[  {\rm x}\, \Phi^{j-;+-}({\rm x},\k) + {\rm x}\, \Phi^{+-;j-}({\rm x},\k)\bigg]
\Bigg\}\Bigg|_{{\rm x}=0}
+3\, \mathcal{F}\textrm{ terms}
\, ,
\label{X_sec_gen_form_Phi}
\end{align}
where ${\cal C}^{ij}_{L}$ and ${\cal C}^{j}_{L}$ are the hard factors which shall be read off from the expressions (\ref{X_sec_L_Fperpmin_Fperpmin}-\ref{X_sec_L_Fplusmin_Fperpmin}).

The leading operator structure $\Phi^{j -;i-}$ is contained in the second line of Eq.~\eqref{X_sec_gen_form_Phi}. The coefficient ${\cal C}^{ij}_{L}$ contains NLP (kinematical twist) corrections and the next-to-eikonal corrections are contained in the $[1+\dots]$ bracket. As already pointed out previously, the two types of the corrections factorize from each other. The two last lines represent new operator structures that are purely next-to-eikonal corrections. The $3\, \mathcal{F}$ term contains eikonal next-to-leading power (genuine twist) corrections and other contributions that are both next-to-leading-power and subeikonal. The analysis of this term is done elsewhere, see Ref.~\cite{Altinoluk:2026qtt}.

The field strength correlators of the form \eqref{def_Phi_correlator_Fmunu_Frhosigma} are usually parametrized in terms of gluon TMD distributions, which are real scalar functions. The parametrization takes the following form
%
\begin{align}
 \Phi^{j -;i-}({\rm x},\k) =&\,
\frac{\delta^{ij}}{2}\, f_1^g({\rm x},\k) +\left[\k^i\k^j-\frac{\k^2}{2}\, \delta^{ij}\right]\frac{1}{2 M^2}\, h_1^{\perp g}({\rm x},\k) \, ,
\nn\\
\Phi^{j-;+-}({\rm x},\k) =&\,
\frac{\k^j}{P_{{tar}}^-}\, \left[f^{\perp g}({\rm x},\k)-i  \bar{f}^{\perp g}({\rm x},\k)\right] \, ,
\nn \\
\Phi^{+-;j-}({\rm x},\k) =&\,
\frac{\k^j}{P_{{tar}}^-}\, \left[f^{\perp g}({\rm x},\k)+i  \bar{f}^{\perp g}({\rm x},\k)\right]
\label{param_Phi}
\, ,
\end{align}
where $M$ is the target mass.
Using the relations \eqref{param_Phi}, we can transform  Eq.~\eqref{X_sec_gen_form_Phi} to the following form
%
\begin{align}
\frac{d\sigma_{\gamma^{*}_{L}\rightarrow q_1\bar q_2}}{dz_1\, d^2\P\, d^2\k}\Bigg|_{{\rm Eik }+{\rm NEik}}
= &\,
  \alpha_{\rm em} e_f^2\, \alpha_{s} \; 
\Bigg\{
{\cal C}^{ f_1^g}_{L}(z_1,\P,\k) 
 \bigg[ 1 +\frac{[\P^2\!+\!\bar Q^2]}{z_1 z_2 W^2} \, \partial_{{\rm x}}\bigg]  
 \bigg[ {\rm x}\, f_1^g({\rm x},\k) \bigg] 
 \nn\\
 &\,
 + {\cal C}^{ h_1^{\perp g}}_{L}(z_1,\P,\k)
 \bigg[ 1 +\frac{[\P^2\!+\!\bar Q^2]}{z_1 z_2 W^2} \, \partial_{{\rm x}}\bigg]  
 \bigg[ {\rm x}\, h_1^{\perp g}({\rm x},\k) \bigg] 
 \nn\\
 &\,
+\frac{\k\!\cdot\!\P}{W^2}\, {\cal C}^{ f^{\perp g}}_{L}(z_1,\P)\,  {\rm x}\, f^{\perp g}({\rm x},\k)
\Bigg\}\Bigg|_{{\rm x}=0}
+3\, \mathcal{F}\textrm{ terms}
\, ,
\label{X_sec_gen_form_scal_TMDs}
\end{align}
where the coefficients are given by
%
\begin{align}
 {\cal C}^{ f_1^g}_{L}(z_1,\P,\k) 
\equiv &\,
\frac{8Q^2\, z_1^2 z_2^2}{[\P^2\!+\!\bar Q^2]^4} 
\left\{\P^2 +(z_2\!-\!z_1)(\k\!\cdot\!\P) \left[-1
+\frac{4\P^2}{[\P^2\!+\!\bar Q^2]}\right]
\right\} \, ,
\nn\\
 {\cal C}^{ h_1^{\perp g}}_{L}(z_1,\P,\k) 
\equiv &\,
 \frac{4Q^2\, z_1^2 z_2^2}{[\P^2\!+\!\bar Q^2]^4}\, \frac{\k^2}{M^2} 
\Bigg\{\left(\frac{2(\k\!\cdot\!\P)^2}{\k^2\P^2}-1\right)\P^2 
\nn \\
&
+(z_2\!-\!z_1)(\k\!\cdot\!\P) \left[-1
+\frac{4\P^2}{[\P^2\!+\!\bar Q^2]}\left(\frac{2(\k\!\cdot\!\P)^2}{\k^2\P^2}-1\right)\right]
\Bigg\} \, ,
\nn\\
{\cal C}^{ f^{\perp g}}_{L}(z_1,\P) 
\equiv &\,
 -32Q^2\, z_1 z_2\,
 \frac{(z_2\!-\!z_1)[\P^2\!+\!m^2]}{ [\P^2\!+\!\bar Q^2]^4} \, .
\label{scal_coeff_contract_L}
\end{align}

In Eq.~\eqref{X_sec_gen_form_scal_TMDs}, the terms in the first two line correspond to leading-power contributions (and NLP terms appear only in the hard factors). They contain both eikonal and subeikonal terms (suppressed by $1/W^2$). The eikonal terms correspond to twist-2 gluon TMDs at $x=0$ due to the absence of the phase factor (see Eq.~\eqref{def_Phi_correlator_Fmunu_Frhosigma}). On the other hand, the next-to-eikonal terms in the first two lines of Eq.~\eqref{X_sec_gen_form_scal_TMDs} can be treated as the first corrections in the Taylor expansion of the twist-2 gluon TMDs around $x=0$. Therefore, Eq.~\eqref{X_sec_gen_form_scal_TMDs} can be rewritten as
%
\begin{align}
\frac{d\sigma_{\gamma^{*}_{L}\rightarrow q_1\bar q_2}}{dz_1\, d^2\P\, d^2\k}\Bigg|_{{\rm Eik }+{\rm NEik}}
= &\,
  \alpha_{\rm em} e_f^2\, \alpha_{s} \; 
\Bigg\{
{\cal C}^{ f_1^g}_{L}(z_1,\P,\k)\:
 {\rm x}\, f_1^g({\rm x},\k) 
 + {\cal C}^{ h_1^{\perp g}}_{L}(z_1,\P,\k)\:
{\rm x}\, h_1^{\perp g}({\rm x},\k)  
 \nn\\
 &
+\frac{\k\!\cdot\!\P}{W^2}\, {\cal C}^{ f^{\perp g}}_{L}(z_1,\P)\:  {\rm x}\, f^{\perp g}({\rm x},\k)
\Bigg \} 
+3\, \mathcal{F}\textrm{ terms} \, ,
\label{X_sec_gen_form_scal_TMDs_x_non_zero}
\end{align}
where the freedom to exponentiate the relevant terms into the x dependent phases was used (within the same accuracy in our beyond-eikonal expansion). 


\section{Summary}

To conclude, the interplay between CGC formalism beyond eikonal accuracy and TMD formalism was studied. Studying DIS dijet production in the correlation limit allowed to find a connection between these two formalisms. The main outcome of these studies is that by resumming a subset of power corrections beyond the eikonal approximation allowed to recover the dependence on $x$ characteristic of TMDs (see Eq.~(\ref{X_sec_gen_form_scal_TMDs_x_non_zero})) in the high-energy formalism.

\newpage

\section*{Acknowledgements}

TA is supported in part by the National Science Centre (Poland) under the research Grant No. 2023/50/E/ST2/00133 (SONATA BIS 13). GB is supported in part by the National Science Centre (Poland) under the research Grant No. 2020/38/E/ST2/00122 (SONATA BIS 10). AC is supported in part by the National Science Centre (Poland) under the research Grant No. 2021/43/D/ST2/01154 (SONATA 17).

\end{document}